\input{psfig}

\documentstyle[prl,aps,twocolumn]{revtex}

\catcode`\@=11

\def\maketitle2{\par 
\begingroup
\let\cite\@bylinecite
\def\thefootnote{\fnsymbol{footnote}}%
\twocolumn[\@maketitle2\vskip2pc]%
\thispagestyle{plain}\@thanks
\endgroup
\def\thefootnote{\arabic{footnote}}%
\setcounter{footnote}{0}%
\let\maketitle2\relax \let\@maketitle2\relax
\let\@thanks\relax \let\@authoraddress\relax \let\@title\relax
\let\@date\relax \let\thanks\relax \let\@abstract\relax 
\let\@pacs\relax}

\def\abstract#1{\gdef\@abstract{{\par 
\bgroup
\ifdim\prevdepth=-1000pt \prevdepth0pt\fi
\hsize\columnwidth
\dimen0=-\prevdepth \advance\dimen0 by17.5pt \nointerlineskip
\small\vrule width 0pt height\dimen0 \relax}{~~}#1\egroup}}

\def\pacs#1{\gdef\@pacs{{\par 
\bgroup
\hsize\columnwidth \parindent0pt
\ifdim\prevdepth=-1000pt \prevdepth0pt\fi
\dimen0=-\prevdepth \advance\dimen0 by20pt\nointerlineskip
\egroup} PACS numbers:~#1}}

\def\@maketitle2{
\@preprint
\@title
\ifdim\prevdepth=-1000pt \prevdepth0pt\fi
\@authoraddress
\@date
\begin{list}{}{\leftmargin=0.10753\textwidth \rightmargin=\leftmargin
\itemsep=1pc\partopsep=-1pc}
\item\@abstract
\item\@pacs
\end{list}
}

\catcode`\@=12

\begin{document}

\title{\vspace*{-0.5cm}
\hspace*{\fill}{\normalsize LA-UR-98-3412} \\[1.5ex]
Various Models for Pion Probability\\
Distributions from Heavy-Ion Collisions}
\author{
A.Z. Mekjian${}^{1,2}$\thanks{E. Mail: mekjian@ruthep.rutgers.edu}{\ }, 
B.R. Schlei${}^{2,3}$\thanks{E. Mail: schlei@LANL.gov}{\ } and
D. Strottman${}^2$\thanks{E. Mail: dds@LANL.gov}{\ }
\\[1.5ex]
{\it 
${}^1$Rutgers University, Department of Physics, Piscataway, NJ 08854, USA\\
${}^2$Theoretical Division, DDT-DO, Los Alamos National Laboratory,
Los Alamos, NM 87545, USA\\
${}^3$Physics Division, P-25, Los Alamos National Laboratory,
Los Alamos, NM 87545, USA
}
}
\date{July 29, 1998}

\abstract{
Various models for pion multiplicity distributions produced in relativistic
heavy ion
collisions are discussed.  The models include a relativistic hydrodynamic
model, a thermodynamic description, an emitting source pion laser
model, and a description which generates a negative binomial description.
The approach
developed can be used to discuss other cases which will be mentioned.  The pion
probability distributions for these various cases are compared.  Comparison
of the pion
laser model and Bose-Einstein condensation in a laser trap and with the
thermal model are made.  The thermal model and hydrodynamic model are also
used to illustrate why the number of pions never diverges and why the
Bose-Einstein correction effects are relatively small.  The pion emission
strength
$\eta$ of a Poisson emitter and a critical density $\eta_c$ are connected in a
thermal model by $\eta/n_c = e^{-m/T} < 1$, and this fact reduces any
Bose-Einstein correction effects in the number and number fluctuation of
pions.  Fluctuations can be much larger than Poisson in the pion laser
model and
for a negative binomial description.  The clan representation of the negative
binomial distribution due to Van~Hove and Giovannini is discussed using the
present description.  Applications to CERN/NA44 and CERN/NA49 data are 
discussed in terms of the relativistic hydrodynamic model. }
\maketitle2
\narrowtext

\section{INTRODUCTION}

\nobreak
Pion multiplicity distributions produced in heavy ion collisions are of
current interest for several reasons. First, pions are by far the main 
component of the produced particles in very high energy heavy ion collisions.
For example, in the SPS experiments at CERN hundreds of pions are produced, 
and this number may go up considerably at RHIC energies.
In the Landau hydrodynamic model, the number of pions scales at $S^{1/4}$
where $S = E_{\rm cm}^2$.  Secondly, pions are a useful tool for studying HBT 
effects coming from Bose-Einstein correlations.  HBT two-particle correlation
experiments give information about source parameters of the emitting system.  
Thirdly, if many pions are produced by a strongly emitting source at
high enough density, a pion laser may be formed according to one model owing
to Bose-Einstein symmetrization effects.  Pions with a nonzero chemical
potential can also show the phenomena of Bose-Einstein condensation at some
critical density.  Since Bose-Einstein condensation of a finite number of atoms
in a harmonic oscillator or laser trap has recently been seen, the possibility 
of seeing effects associated with the formation of a pion laser or an enhanced
condensation into a ground state would be interesting.  A comparison of
Bose-Einstein condensation of atoms in a harmonic oscillator trap and the pion
laser model will be given in this paper.  Fourthly, the possibility of
intermittency behavior in pion distributions has been a concern for over a
decade.  A distribution that is widely used to discuss intermittency is the
negative binomial distribution, and a model which generates a negative
binomial distribution will be developed in the framework of the present
description.  The clan representation for the negative binomial distribution
due to Van~Hove and Giovannini will also be discussed.  This representation
is a useful way of characterizing experimental data.

In this paper we present a unified description of several currently
discussed models for evaluating pion distributions.  Our main focus will be 
on four particular cases, but other possibilities exist and will be discussed 
briefly.  The examples we will consider in detail
are 1)~a thermodynamic description, 2)~a relativistic hydrodynamic model, 
3)~an emitting source model, and 4)~a model which gives a negative binomial
distribution. These cases cover a wide spectrum of possibilities.  
Thermodynamic descriptions have been extensively used \cite{fer50}--\cite{qm97}
for the production of pions in heavy ion collisions and also are discussed in 
the quark matter meetings \cite{qm96,qm97}. Emitting source models are 
reviewed in Refs. \cite{boa90,wei97}. The emitting source model that we 
discuss in detail is originally due to Pratt \cite{pra93} and discussed 
further in Refs. \cite{gao95,cso77}.
Application of the negative binomial distribution to high energy collision 
data can be found in Refs. \cite{hov86}--\cite{car83}, and its connection 
with intermittency are discussed in Refs. \cite{bia86,bia92}.  Hydrodynamic 
descriptions are also frequently applied to heavy ion collisions, and their 
descriptions appear in the quark matter proceedings, Refs. \cite{qm96,qm97}.  
The relativistic hydrodynamic model gives a good description of single and 
double inclusive data from Cern experiments carried out by the CERN/NA44 and 
CERN/NA49 collaborations.

We will use an approach that initially was developed for cluster distributions
\cite{mek90}--\cite{chapre}, but applications to cycle class problems were 
noted \cite{mek90} and further studied in more detail in Refs. 
\cite{cha96}--\cite{chapre}.  Cycle class problems appear in Bose-Einstein and
Fermi-Dirac statistics because the density matrix must be symmetrized or 
antisymmetrized in Feynman's path integral approach to statistical mechanics
\cite{fey72}.  A detailed application of the approach discussed below 
to Bose-Einstein condensation phenomena can be found in Refs. 
\cite{cha94,chapre}. The connection between cluster models and cycle class 
problems is shown in Fig.~1.  Specifically, a cycle of length $k$ corresponds 
to a cluster of size $k$, and the number of cycles of length $k$ corresponds 
to the number of clusters of size $k$.

Application of results from cluster models to the theory of disordered
systems can be found in Refs. \cite{mek97,cha98}. Some of the results of this 
reference were connected to probability distributions which appear in 
{\it Quantum Optics}.  Methods from {\it Quantum Optics\/} have been used by 
Weiner \cite{and93}--\cite{bernd8} {\it et~al.}\ to discuss particle production and
correlations in heavy ion collisions.

\section{PION PROBABILITY DISTRIBUTIONS}

\nobreak
In this section we will discuss several models of pion production in heavy ion
collisions.  These models, which are a thermodynamic description, a 
relativistic hydrodynamic model, an emitting source model, and a negative
binomial model, can be described using a unified formalism.  Before giving
specific results, we will give some general expressions that we will use.
These expressions were initially developed for the description of the 
fragmentation of nuclei into clusters.  However, because of the correspondence 
between cycles and clusters shown in Fig.~1, the cluster results can be used to
describe cycle class problems, and the pion production models that we consider 
can be re-expressed in such terms.  This correspondence has already been 
illustrated in the description of Bose-Einstein condensation.

This section is divided into various subsections.  First, we give some
general results.
Then we will apply these results to the cases mentioned above.  Comparisons
will be made for the various cases and with other situations already discussed
such as Bose-Einstein condensation.

\subsection{General Results}

\nobreak
In describing a fragmentation process, partitioning problems, or cycle
class problems, a weight is given to each division of $A=\sum_k k n_k$.  
Figure~2 illustrates the possible divisions of $A = 4$.  A vector $\vec n = 
(n_1, n_2, \ldots, n_k, n_A)$ specifies the division where $n_k$ is the number 
of clusters with $k$ nucleons, cycles of length $k$, etc.  A general block
picture for $\vec n$ is shown in Fig.~1, and the specific block pictures of
$A = 4$ are shown in Fig.~2 along with cluster and cycle class divisions and 
harmonic oscillator excitations. The type of weight $W_A (\vec n, \vec x)$ 
given to any $\vec n$ that we will consider can be written as
$$W_A \left(\vec n, \vec x\right) = {1 \over Z_A (\vec x)} \prod_k
{x_k^{n_k} \over n_k!} \eqno (1)$$
Here $n_k!$ are Gibbs factorials, $Z_A (\vec x)$ is the canonical partition
function, and $x_k$ is a function of thermodynamic quantities such as volume 
$V$, temperature $T$.
This $x_k$ was called a tuning parameter in Ref. \cite{mek90}, and a specific 
form $x_k = x/k$ was studied in detail in Refs. \cite{mek90,mek91}.  
Other forms for $x_k$ were used to study percolative features of simple 
statistical model \cite{cha96} and Bose-Einstein condensation of atoms in a 
box \cite{cha94} and atoms in a laser trap \cite{chapre}.  The form of $x_k$ 
for the pion models to be discussed will be given below.

Given $W_A (\vec n, \vec x)$ of the type of Eq.~(1), various quantities are
easily calculated.  For example the ensemble averages of $n_k$, given by
$\left<n_k\right> = \sum W_A (\vec n, \vec x) n_k$, where the sum is over all 
partition of $A$, is simply \cite{mek90,lee92}
$$
\left<n_k\right> = x_k
{Z_{A-k} (\vec x) \over Z_A (\vec x)} \eqno (2)
$$
This result can be easily understood heuristically by noting that the $n_k$
times the Gibbs factorial $1/n_k!$ gives $1/(n_k - 1)!$.  This operation 
removes one column of $k$ blocks from Fig.~1.  The sum over all partitions 
of $A$, which we call $\pi_A(\vec n)$ in $\sum W_A (\vec n, x) n_k$, then 
gives rise to the factor $Z_{A-k}(x)$.  The $x_k$ factor in Eq.~(2) is from 
the shift $x_k^{n_k}= x_k x_k^{n_k - 1}$ where the $x_k^{n_k-1}$ is the 
appropriate power of $x_k$ for the partitions with one column of $k$ blocks 
removed.  The $Z_A (\vec x)$ in Eq.~(2) is present in $W_A (\vec n, \vec x)$ 
and is in the final answer.
Since the constraint $A = \sum_k kn_k$ always applies to any partition of $A$, 
then it is also true when written as $A = \sum_k k\left<n_k\right>$.  
Substituting the result of Eq.~(2) into this constraint equations gives $A =
\sum_k kx_kZ_{A-k}/Z_A$, and rearranging terms results in the recursive
relation
$$
Z_A (\vec x) = {1 \over A} \sum_k kx_k Z_{A-k} (\vec x) \eqno (3)
$$
with $Z_0 = 1$.  This recursive equation is a very simple way of obtaining
$Z_A (\vec x)$, and it applies to any $W_A (\vec n, \vec x)$ of the form of 
Eq.~(1).  To use it, the $x_k$'s
have to be specified.  As an example $x_k = x/k$ gives \cite{mek90,mek91}
$$
Z_A (x) = x (x + 1)\ldots (x+A-1)/A! =
{\Gamma (A + x) \over \Gamma (x) A!} \eqno (4)
$$

\noindent
The $Z_A(\vec x)$ is the canonical partition function of a system of $A$
objects ($A$ nucleons, $A$ atoms, \dots).  The grand canonical partition 
function $z(\vec x, u)$ is  obtained from the canonical partition function 
$Z_A (\vec x)$ or the weight $W_A (\vec n, \vec x)$.  For example,  the 
generating function of $W_A (\vec x) = \sum_{\pi_A (\vec n)} W_A (\vec n,
\vec x)$ is \cite{mek90}
$$
z (\vec x,u) \equiv \exp \left[ux_1 + u^2 x_2 + u^3 x_3 + \ldots\right] =
\sum_{A=0}^\infty W_A (\vec x) u^A \eqno (5)
$$

The $u$ is taken to be $u = e^{\beta\mu}$ in thermodynamics, where $\mu$ is
the chemical potential.  For $\mu = 0$, $u = 1$ and
$$
z (\vec x) = \exp \left[\sum_{k=1}^\infty x_k\right] \eqno (6)
$$
The ratio
$$
Z_A (\vec x)/z(\vec x) = P_A (\vec x) \eqno (7)
$$
is the ratio of the canonical partition function to the grand canonical
partition function.
The mean and variance of $A$ follow once the $x_k$'s are specified:
\begin{eqnarray*}
\left<A\right> &=& \sum_k kx_k &(8)\cr
\noalign{\rm and}
\left<A^2\right> - \left<A\right>^2 &=& \sum_k k^2 x_k\;. &(9)\cr
\end{eqnarray*}
These last two equations follow from Eq.~(5) by differentiating this
equation with respect
to $u$ and setting $u = 1$.  These expressions form the main general
results to be used
in this paper.  Other results can be found in Refs. \cite{mek90}--\cite{cha94}.

\subsection{Pion Distributions in a Thermal Model}

\nobreak
We begin this subsection by first giving some well-known results on the thermal
properties of an ideal gas of relativistic pions.  Some of these results will 
then be used to motivate some ideas, developments, and comparisons to be made 
later. Moreover, since the thermal description is extensively used, many 
readers will recognize these results and this is helpful for the subsequent 
development of the choice of $x_k$ for this particular model and other models 
to follow.

In a thermal model or in a description based on statistical mechanics, the
distributions associated with pions are obtained from the assumption that an 
equilibrium is established in some interaction volume $V$ at some temperature 
$T$. Specifically, statistical thermodynamics gives the Bose-Einstein 
occupancy factor
$$
f(\epsilon) = {1 \over e^{\beta\epsilon}-1} \eqno (10)
$$
for an energy level at $\epsilon$ and $\beta = 1/k_BT$.  This result is
based on the grand canonical ensemble with the pion chemical potential 
$\mu = 0$.  For pions in a box of volume $V$, the thermal properties of pions 
can be obtained from the density of states factor $Vd^3p/h^3$ and from the 
energy $\epsilon$, momentum $\vec p$, relation $\epsilon = (p^2 + m^2)^{1/2}$.
In particular, with $\hbar = c = k_B = 1$, the following relations can be 
obtained for thermal pions:
\begin{eqnarray*}
{N \over V} &=& {1 \over 2\pi^2} m^2 T \sum_{k=1}^\infty {1 \over k} K_2
\left(k {m \over T}\right)\\
{E \over V} &=& {1 \over 2\pi^2} m^3 T \sum_{k=1}^\infty {1 \over k}
\left[{3 \over 4} K_3 \left(k {m \over T}\right) + {1 \over 4} K_1 \left(k
{m \over
T}\right)\right]\\
PV &=& {1 \over 2\pi^2} m^2 T \sum_{k=1}^\infty {1 \over k^2} K_2
\left(k {m \over T}\right) &(11)\cr
{S \over V} &=& \left((E/V) + p\right)\big/T\\
\end {eqnarray*}
Here $N/V$ is the total number density, $E/V$---the energy density,
$pV$---the equation
of state, and $S/V$---the entropy density of pions.  The various $K$'s are
MacDonald functions.  The ultrarelativistic photon-like limit $m \to 0$ or
$T \gg m$ and the nonrelativistic limit $\epsilon = p^2/2m + m$ can be 
obtained from these equations using the following limits
for the MacDonald functions $K_\nu$ that appear.  With $z = km/T:K_\nu (z) =
\left(\Gamma (\nu)/2\right)\big/\left(1/2 z\right)^\nu$ for $z \to 0$ and
$K_\nu (z) = \sqrt{\left(\pi/2z\right)}\, e^{-z}$ for $z \to \infty$.  Here
$\Gamma(\nu)$ is a gamma function.  In the above Eq.~(11), the sum over $k$ 
gives the corrections arising from Bose-Einstein or BE statistics, with the 
$k = 1$ term the Maxwell Boltzmann or MB limit.  Specifically, the $k = 2, 3, 
\ldots \infty$ give the enhancements coming from BE statistics.  For example, 
for $N/V$ the ratio $N/V\big/(N/V)_{\rm MB} = \zeta (3)$ in the massless pion 
limit.  Here $(N/V)_{\rm MB}$ is the $k = 1$ term only.  The $\zeta (3) = 1.2$ 
is the Riemann zeta function.  The results of Eq.~(11) arise when the 
occupancy factor of Eq.~(10) is expanded as a power series $f(\epsilon) = 
e^{-\beta\epsilon} \left(1+e^{-\beta\epsilon} + e^{-2\beta\epsilon} + 
\ldots\right)$.  The integrals over $(Vd^3p/h^3)e^{-k\beta\epsilon}$ with 
$\epsilon = \sqrt{p^2 + m^2}$ give rise to the various $K$'s.

The probability of having $n_k$ pions in level $k$, given that the mean
number in that level is $(n_k) = 1/\left(e^{\beta\epsilon_k}-1 \right)$ is 
the geometric distribution \cite{mor69}
$$
p(n_k) = {\left<n_k\right>^{n_k} \over \left(\left<n_k\right> +
1\right)^{n_k + 1}}
\eqno (12)
$$
The $\left<n_k\right> = \sum\limits_{n_k = 0}^\infty p(n_k)n_k$ and
$\left<n_k^2\right> - \left<n_k\right>^2$ is
$$
\left<n_k^2\right> - \left<n_k\right>^2 = \left<n_k\right> \left(1 +
\left<n_k\right>\right)
\eqno (13)
$$
The Poisson limits is $\left<n_k^2\right> - \left<n_k\right>^2 =
\left<n_k\right>$ and is
realized when MB statistics applies.  The above expression on fluctuations
pertain to a particular energy level.  We now turn to an evaluation of the
fluctuations in the total number of pions and its associated probability
distribution.  Such quantities are important in an event-by-event analysis of
data.  The probability of $N$-pions is given by the ratio of the canonical
partition function, $Z_N$,  to the grand canonical partition function $z$,
with $z
=\sum\limits_{N=0}^\infty Z_N$ and $Z_O = 1$.  To obtain $Z_N, z$ we use
results from Sec.~II.A with $u = e^{\beta\mu} = 1$.  For a thermal model the
choice of $x_k$ is
$$
x_k = {VT^3 \over 2\pi^2} \left({m \over T}\right)^2 {1 \over k^2} K_2
\left(k {m
\over T}\right)
\eqno (14)
$$
Later, we will give other choice of $x_k$.  That this is the correct choice
follows from $\left<N\right> = \sum kx_k$ where $\left<N\right>/V$ in the 
grand canonical ensemble is given by Eq.~(11).  Also $z = e^{PV/T} = 
e^{\Sigma_k x_k}$ where $PV/T = \sum\limits_k x_k$ follows also from Eq.~(11), 
and $z = e^{\Sigma_k x_k}$ was a result developed in Sec.~II.A.  The result for 
$z = e^{\Sigma x_k}$ can be expanded to give the $W_N (\vec x)$ of Eq.~(5) and 
its associated $W_N (\vec n, \vec x) = \prod x_k^{n_k}/n_k!\big/Z_N (\vec x)$ 
of Eq.~(1).  The $W_N (\vec n, \vec x)$ is the weight given to $\vec n =
\left(n_1, n_2 \ldots n_A\right)$ in a cycle class decomposition of a
permutation.  For example the permutation
$$
\pmatrix{1&2&3&4&5&6&7\cr
1&3&2&4&7&5&6} \eqno (15)
$$
has two cycles of length~1 which are $1 \to 1$ and $4 \to 4$, one cycle of
length~2 which is $2 \to 3 \to 2$, and one cycle of length~3 which is 
$5 \to 7 \to 6 \to 5$.  The cycles appear in Feynman's path integral density 
matrix approach in {\it Statistical Mechanics} \cite{fey72}, and permutations 
appear because the density matrix must be symmetrized.  An example of this 
approach to the ideal Bose gas in a box \cite{cha94} and to the ideal Bose gas 
in a harmonic oscillator trap \cite{chapre} shows the utility of the 
recurrence relation of Eq.~(3) in obtaining the canonical partition function 
$Z_N(\vec x)$ and all its associated thermodynamic quantities.  We will briefly
summarize these two cases in the next section.

Using this choice of $x_k$, the probability of $N$ pions is $P_N = Z_N/z$ where
$Z_N$ is obtained from Eq.~(3) with $x_k$ given by Eq.~(14).  The fluctuations
$\left<N^2\right> - \left<N\right>^2 = \sum k^2 x_k$.  An alternative method 
for obtaining $P_N$ was recently given by Becattini, Giovannini, and Lupia
\cite{bec96}.  They used projection methods to obtain $P_N$ which involves 
taking $N$-derivatives of $z$ with respect to $u$ and setting $u = 0$ .
Specifically
$$
P_N = {1 \over N!} \lim_{w\to 0} {d^N \over dw^N} \exp
\left[- {V \over 2\pi^2} \int d^3 p \log \left(1 -
e^{-\epsilon/T}\omega\right)\right]
\eqno (16)
$$
Table~1 summarizes $P_N$ for $T = 163$~MeV, $V = 22.5$, and $m/T = 0.85$.
This choice generates the same $P_N$ distribution given in Ref. \cite{sta92}.  
Here the $P_N$'s are calculated by the simple recurrence relation of Eq.~(3).  
Computationally, Eq.~(3) is very easy to work with since computers evaluate 
recursive relations quickly as well as the sums for $\left<N\right>$ and 
$\left<N^2\right> - \left<N\right>^2$ that appear in Eq.~(8).

Two limiting behavior of $x_k$ are of interest.  In the ultrarelativistic
limit or massless pion limit, the $x_k$ of Eq.~(14) is simply
$$
x_k = {V \over \pi^2} T^3 {1 \over k^4} \eqno (17)
$$

and in the nonrelativistic limit
$$
x_k = {V \over \lambda_T^3} e^{-k m/T} {1 \over k^{5/2}} \eqno (18)
$$
where $\lambda_T = h/(2\pi\, mT)^{1/2}$.  The factor $e^{-k\,m/T}$ is the
Boltzmann factor cost function for creating $k$-pions, and this factor arises
from the mass of the pion.  We will introduce a quantity $\tau$, where $x_k
\sim 1/k^\tau$.  In cluster models $\tau$ is Fisher's critical exponent 
\cite{sta92}.  In Eq.~(17) $\tau = 4$ and in Eq.~(18) $\tau = 5/2$.

\subsection{Relativistic Hydrodynamic Model and Applications to CERN/NA44 and 
CERN/NA49 Experiments}

\nobreak
We also calculated the pion probability distribution using a relativistic
hydrodynamic model: HYLANDER-C \cite{bernd15}. This particular model belongs 
to the class of models which apply 3+1-dimensional relativistic 
one-fluid-dynamics. It provides fully three-dimensional solutions of the 
hydrodynamical relativistic Euler-equations \cite{euler}. HYLANDER-C and its 
earlier version, HYLANDER \cite{udo}, have been successfully applied to 
various heavy-ion reactions at SPS energies. HYLANDER-C and HYLANDER were used 
to reproduce \cite{jan} simultaneously mesonic and baryonic rapidity and 
transverse momentum spectra of the $S+S$ reaction at 200 $AGeV$. Corresponding 
measurements have been performed by the NA35 Collaboration \cite{wenig}. Based 
on the successful description of the measured single-inclusive spectra, 
predictions for Bose-Einstein correlation (BEC) functions \cite{bernd2,bernd3}
were made. Those predictions turned out to agree quantitatively with the 
measurements \cite{alber,QM95}. The model also reproduces the photon data for 
$S+Au$ collisions at SPS energies \cite{axel} and gives a simple explanation
for the ``soft-$p_{\perp }$ puzzle'' \cite{udo91} and the complex behaviour of
the radii  extracted from pion and kaon correlations and explains the 
difference in the extracted  radii for pions and kaons in terms of a cloud of 
pions due to the decay of resonances which surrounds the fireball (pion halo) 
\cite{bernd2}. 

In the following, we discuss some further results \cite{bernd_eos} for 
158A~GeV Pb + Pb collisions measured by the CERN/NA44 
\cite{NA44xu,NA44baerden} and CERN/NA49 Collaborations \cite{NA49jones}.  
The by HYLANDER-C reproduced Pb+Pb data have been obtained while using an 
equation of state (EOS) with a phase transition to a quark-gluon plasma at a 
critical temperature $T_C=200\:MeV$ ({\it cf.} Refs. \cite{redlich}, 
\cite{bernd8} and Refs. therein). This EOS does not depend on the baryon 
density, and thus the freeze-out energy density translates directly into a 
fixed freeze-out temperature $T_f$. The choice for the freeze-out temperature 
is in these calculations, $T_f=139 MeV$.

The number of thermal pions ($\pi^+$, $\pi^-$, and $\pi^0$) that are produced 
in the model calculations is 853. In particular, this number does not include 
any resonance decay contributions. But this total number of thermal pions 
includes the Bose-Einstein enhancement. To be specific, the thermal pion 
spectra are calculated by evaluating the integral

$$
E\frac{d N}{d^3 p}\:=\:
\frac{1}{(2\pi)^3}\:
\int_{\Sigma}
\frac{p^\mu d\sigma_\mu(x'_\mu)}
{\exp \displaystyle{\left[ \frac{p^\mu u_\mu(x'_\mu)}
{T_f(x'_\mu)} \right] - 1}}\:,
\eqno (19)
$$

where $d\sigma^\mu$ is the differential volume element and the integration is 
performed over the freeze-out hypersurface, $\Sigma$. $u^\mu(x)$ and $T_f$ are 
the four velocity of the fluid element at point $x$ and the freeze-out 
temperature, respectively. $p^\mu = (E, \vec{p}\,)$ is the four-momentum of a 
pion.

By expanding the Bose-Einstein occupation factor in terms of a 
Maxwell-Boltzmann factor and a series of higher order corrections, we can 
obtain the distribution of $x_k$'s which are summarized in Table~2.

Using these results for $x_k$, the fluctuation in the thermal pions is
$\left<N^2\right> - \left<N\right>^2 = \sum k^2 x_k = 312.97$ or a 20\%
enhancement over the Poisson result of $x_1 = 260$. The probability distribution 
associated with the thermal pions is obtained from $P_N = Z_n/z$ where $Z_N$ is 
given by Eq.~(3) and $z$ is given by Eq.~(6).  Figure~3 shows the result for 
$P_N$, which is a perfect Gaussian distribution (solid line) with mean 
$\langle N \rangle = 283.9$ and variance $\langle N^2 \rangle -
\langle N \rangle^2 = 312.97$.

\subsection{Some Special Examples}

\nobreak
In this subsection, we will illustrate some of the general results of Sec.~II.A
with three specific examples.  For massless particles in a box of sides $L$ in
$d$-dimensions (neglecting spin), the $x_k$ is
$$
x_k = {L^d \over 2^{d-3}\pi^{d-1}} \left({k_BT \over \hbar c}\right)^d
{1 \over k^\tau} \equiv {x_d \over k^\tau} \eqno (20)
$$
where $\tau = d+1$.  Then $z = \exp [x_d \zeta (\tau)]$, $\left<N\right> = 
x_d \zeta (d)$ and $\left<N^2\right> - \left<N\right>^2 =
x_d \zeta (d-1)$ with $\zeta (n) = \sum_{k=1}^\infty 1/k^n$.  Some values of 
$\zeta (n)$ that often appeal in statistical mechanics are
$\zeta (1) = \infty$, $\zeta (2) = \pi^2/6 = 1.645$, $\zeta (3) = 1.202$,
$\zeta (4) = \pi^2/90 = 1.082$, $\zeta (3/2) = 2.612$.  In two dimensions 
$\left<N^2\right> - \left<N\right>^2 = \infty$ so that  the fluctuations of 
massless particles becomes large.

For nonrelativistic atoms in a box \cite{cha94} in $d$-dimensions
$$
x_k = {x \over k^\tau} + {1 \over k} \eqno (21)
$$
with $\tau = d/2 + 1$, $x = L^d/\lambda_T^d$, $\lambda_T = h/(2\pi mT)^{1/2}$.
Condensation occurs in $d = 3$ at $x = x_c = A/\zeta (3/2) = 
V/\lambda_{T_c}^3$ in the limit $A$, $V \to \infty$.  Manifestation of this 
Bose-Einstein condensation are 1)~a power law distribution of cycle lengths
\cite{cha96,cha94}: $\left<n_k\right> \sim 1/k^{5/2}$, 2)~the
specific heat $C_V$ has a cusp \cite{mor69}, and 3)~the ground state occupany
$n_{G.S.}$ becomes macroscopically large \cite{mor69}: $n_{G.S.}/A = 1 - x/x_c
= 1 - (T/T_c)^{3/2}$, for $T \leq T_c$.  The $T_c$ is determined by the 
density $A/V$ from $A/\zeta (3/2) = V/\lambda_{T_c}^3$.  For $d=1,2$ $T_c 
\to 0$ so that there is no sudden condensation at a nonzero $T$.  The 
occupancy of any level with energy $\epsilon_k$ is
$$
n_k = \sum_{k=1}^A e^{-\beta\epsilon_k k} {Z_{A - k} \over Z_A} \eqno (22)
$$
For $A \to \infty$, $Z_{A - k}/Z_A \to e^{\beta \mu k}$ and thus we obtain the
well-known occupancy factor
$$
n_k = {1 \over e^{\beta(\epsilon_k-\mu)}-1}\;. \eqno (23)
$$
As $T \to T_c$, $\mu \to \epsilon_0 \equiv 0$.

For atoms in a laser trap given by a harmonic oscillator well with
frequency $\omega$ and level spacing $\hbar\omega$ \cite{chapre}
$$
x_k = {1 \over k}\, {x^{kd/2} \over (1 - x^k)^d} \eqno (24)
$$
where $x = e^{-\hbar\omega/k_BT}$.  For small $x$
$$
x_k = {x^{kd/2} \over k^{d+1}} \left({k_BT \over \hbar\omega}\right)^d
$$
There is no condensation at nonzero $T$ for $d=1$, but there is
condensation for $d = 2,3$ at $T_c \not= 0$.

\subsection{Emitting Source, Pion Laser Model}

\nobreak
A simple and exactly solvable emitting source models for pions was first
introduced by S.~Pratt \cite{pra93}.  The model has an interesting feature.
Namely, a Poisson emitter of pions can behave like a pion laser when 
Bose-Einstein symmetrization effects are included.  A manifestation of this 
behavior is a large enhancement in the number of pions.  This enhancement is 
into a zero momentum state and is similar to Bose-Einstein condensation of 
atoms in a box which also condense into the zero momentum state.  An important 
result of this condensation is a reduction in the intercept in the two particle
correlation functions.  Such effects were discussed in Refs. 
\cite{and93}--\cite{mor69}.  Recently Cs\"org\"o and Zim\'any \cite{cso77} 
have given simple exact expressions for Pratt's model.
Some simple results were also presented by Chao, Gao, and Zhang \cite{gao95}.
This subsection presents some further discussions of this model using results
from Sec.~II.A when applied to this case and shows some similarities with the
Bose-Einstein condensation of atoms in a laser trap as given by Eq.~(24).

First, we will give expressions for $x_k$.  Once $x_k$ is given, then all the
general results of Sec.~II.B apply.  In Pratt's model, $\eta$ is the mean pion
multiplicity generated by a Poisson emitter.  Then, in the Poisson limit, the
probability of having $N$ pions is $P_N = \left(\eta^N/N!\right)e^{-\eta}$.  
This is also the result of the thermal model of Sec.~II.B in the 
Maxwell-Boltzmann limit, where $\eta = x_1 = \left(VT^3/2\pi^2\right) (m/T) 
K_2 (m/T)$.  Corrections to this result obtained from Bose-Einstein 
symmetrization effects can be developed by evaluating a quantity $x_k$ (called 
$C_k$ in Ref. \cite{pra93}).   The $C_k$ are referred to as combinants
\cite{cso77}.  
The importance of combinants can be found in the work of
Gyulassy and Kaufman \cite{gyu78,kau78}.   While the $x_k$ of the thermal 
model is given by Eq.~(14), in the emitting source model $x_k$ is given by 
evaluating

$$
x_k = {\eta^k \over k} \int \prod_{i=1}^k  d^3 \vec x_i d^3 \vec p
\prod_{i=1}^{k-1} d^3 p_i
e^{i\Sigma (\vec p_{i-1} - \vec p_i) \cdot \vec x_i}
$$
$$
\times\prod_{i=1}^k s \left({1 \over 2} \left(p_{i-1} + p_i\right)\;,\; x_i\right)
\eqno (25)
$$
where $p_0 = p = p_k$ and $x_1 = \eta$.  We neglect any temporal dependence
to keep the discussion simple.  

The $s(\vec p, \vec x)$ is the source strength.
As in Ref. \cite{pra93} this source strength is taken to be a thermal Gaussian 
of the form
$$
S(\vec p, \vec x) = {1 \over \left((2\pi)^2 R^2 mT\right)^{3/2}}
e^{-\vec P^2/2mT + \vec x^2/2R^2} \eqno (26)
$$
For this choice of $s(\vec p, \vec x)$, the evaluation of the $x_k$'s can
be done
analytically \cite{gao95}.  We now give a discussion of the liming behavior of
$x_k$ before giving the general behavior.  For small $k$ and $R^2/2 \gg 1/8mT$
(typically $R^2/2 \sim 6$ and $1/8mT \sim 1/6$)
$$
x_k = \left(R^2 mT\right)^{3/2} \left({\eta \over
\left(R^2mT\right)^{3/2}}\right)^k {1 \over k^4} \eqno (27)
$$
Thus $x_k \sim 1/k^4$ and $\tau = 4$ for small $k$.  For large $k$
$$
x_k = {1 \over k} \left({\eta \over \eta_c}\right)^k
\eqno (28)
$$
Now $x_k \sim 1/k$ and $\tau = 1$.  The $\eta_c$ is given by
$$
\eta_c = \left(R^2 mT + \sqrt{R^2 mT} + {1 \over 4}\right)^{3/2}
\eqno (29)
$$
Table~3 gives some values of $x_k$ for $R^2/2 = 6$ and $1/8 mT = 1/6$.

The general behavior for $x_k$ was explicitly written by Cs\"org\"o and
Zim\'any \cite{cso77}, and is
$$
x_k = {\eta^k \over k} {1 \over \left(\gamma_+^{k/2} - \gamma_-^{k/2}\right)^3}
\eqno (30)
$$
with $\gamma_\pm = 1/2 \left(1 + y \pm \sqrt{1 + y}\right)$ and
$y = 2R^2 mT - 1/2$.  The $\eta_c = \gamma_+^{3/2}$ in this notation.  We
rewrite this as
$$
x_k = \left({\eta\over \eta_c}\right)^k
{1 \over k\left(1 - \left(\left(\gamma_- \over
\gamma_+\right)^{1/2}\right)^k\right)^3} \eqno (31)
$$
to illustrate its close relation to atoms in a laser trap as given by
Eq.~(24).  For
$d = 3$, Eq.~(24) gives
$$
x_k = {x^{3k/2} \over k \left(1 - x^k\right)^3} \eqno (32)
$$
with $x = e^{-\hbar \omega/k_B T}$.

Once the $x_k$'s are given, the $z = \exp [\sum x_k]$, $\left<N\right> =
\sum kx_k$, $\left<N^2\right> -\left<N\right>^2 = \sum k^2 x_k$ and
$P_N = z_N/z$ follow with $Z_N$ given by the recurrence expression of
Eq.~(3).  Since $x_k = 1/k (\eta/\eta_c)^k$ for large $k$, $\left<N\right> \to
\infty$ when $\eta \geq \eta_c$.  At $\eta = \eta_c$ the $\left<N\right> \sim
\sum 1/k$ which diverges logarithmically.  We now compare these results with
thermal models.  In thermal models, $x_k$ is given by Eq.~(14) and its
nonrelativistic and ultrarelativistic limits in Eq.~(17) and Eq.~(18).  No
divergence in $\left<N\right>$ occurs in either the ultrarelativistic limit
where $\left<N\right> = \left(V/\pi^2\right) T^3 \zeta (3)$, or the 
nonrelativistic limit where $\left<N\right> = \left(V/\lambda_T^3\right) 
g_{3/2} \left(m/T\right)$  with $g_{3/2} (m/T) = \sum 
\left(e^{-m/T}\right)^k\big/k^{3/2}$.  Note, however, that fluctuations,
$\left<N^2\right> - \left<N\right>^2$, in the two-dimensional ultrarelativistic
case discussed in Sec.~II.C can be infinite.  To further illustrate the
difference, we note that in the thermal model all $x_k$'s are given by 
Eq.~(14) so that $x_1 \equiv \eta$ is not independent of $V, T,m$ where $V 
\sim R^3$.  Table~4 shows values of $x_k$ for $V = {4\pi \over 3} R^3$ with 
$R^2/2 = 6$ and $mT = 3/4$.
Using $\eta = x_1$, the $x_k$ can be rewritten as $x_k \sim \eta^k$ and the
results are also shown when written in this form.  The results of Table~4 are
very similar to Table~3.  However, the large $x_k$ limit is
$\left(V/\left(\lambda_T^3 \cdot k^{5/2}\right)\right) e^{-k m/T}$.  If we take
$x_1$ to be this result with $k = 1$, the $x_k =
\left(1/k^{5/2}\right)\,\left(\eta/\eta_c\right)^k$ with $\eta_c =
V/\lambda_T^3 \sim \left(R^2 mT\right)^{3/2}$, which is similar to Eq.~(29).

\noindent
Note that $\eta = \eta_c e^{-m/T}$  so that $\eta/\eta_c = e^{-m/T} <
1$.  Consequently, the sum $\sum x_k k = \left<N\right>$ always converges.
Thus
there can be no result $\left<N\right> \to \infty$.  The Boltzmann factor
$e^{-m/T}$ suppresses the property $\left<N\right> \to \infty$.  This hidden
connection between $\eta$, $\eta_c$, and $\eta/\eta_c < 1$ is not present in
the pion laser model which treats $\eta$ as independent of $\eta_c$ or $x_1$ as
independent of the other $x_k$'s.  The thermal model has this
connection and shows why no divergence in $\left<N\right>$ ever occurs in
the thermal model.  Similar conclusions also apply to the relativistic
hydrodynamic model.

\subsection{Negative Binomial Distribution}

\nobreak
As we have seen, pion probability distributions are determined once the
choice of the $x_k$'s is specified.  The thermal equilibrium model specifies a
particular form for $x_k$ as does the emitting source model.  Bose atoms in a
box or laser trap also can be discussed in terms of $x_k$.  In this
subsection we will present a choice for $x_k$ which results in a negative 
binomial distribution.
The choice given below is partly motivated by results from the other models.
The negative binomial is a probability distribution that is frequently used
to characterize pion yields \cite{hov86}--\cite{hov89}.  It has also been 
used in the discussion of intermittency \cite{bia86,bia92}.  Van~Hove and 
Giovannini \cite{hov86}--\cite{hov89} has given a clan model representation 
of it which will also be mentioned.

The various forms for $x_k$ that we have encountered so far in this paper and
also in the application of this approach to cluster yields suggest looking at 
a behavior for $x_k$ given by
$$
x_k = ay^k/k^\tau \eqno (33)
$$

For example, the nonrelativistic thermal model has $a = V/\lambda_T^3$, $y =
e^{-m/T}$, and $\tau = 5/2$.  The ultrarelativistic thermal model has $a = {V
\over \pi^2} T^3$, $y = 1$, and $\tau = 4$.  The emitting source model, for
small $k$, is $a = \left(R^2 mT\right)^{3/2}$, $y = \eta/(R^2mT)^{3/2}$, 
$\tau = 4$ and is $a = 1$, $y = \eta/\eta_c$, $\tau = 1$ for large $k$.  We 
will now show that the choice
$$
x_k = ay^k/k\;, \eqno (34)
$$

\noindent
which has $\tau = 1$, generates a negative binomial probability distribution.
This way of generating a negative binomial distribution can be found in
Ref. \cite{que49}, more recently in Ref. \cite{heg93}, and was used in 
Refs. \cite{mek90} and \cite{lee92}.  The $a$ and $y$ will be related to the 
mean number of pions and to their fluctuations.  We will also connect them to 
the clan variables of Van~Hove.  Using results from Ref. \cite{mek90} and 
Sec.~II.A, the following behavior for various relevant quantities are
obtained.  First, the grand canonical partition function $z = \exp
\left(\sum x_k\right) = 1/(1-y)^a$ which can easily be verified by noting
$\sum\limits_k ay^k/k = -a \log (1-y)$.
The canonical partition function $Z_n$ can be obtained and is \cite{mek90} 
$Z_N = y^N \Gamma (a + N)/\Gamma (a) N!$---see also Eq.~(4).  The
extra $y^N$ that appears here is from the $y^k$ in $x_k$.  Note that
$\prod\limits_k (y^k)^{n_k} = y^N$ since $\sum k n_k = N$, where the
$\prod\limits_k (y^k)^{n_k}$ appears in $W_A (\vec n, \vec x)$ of Eq.~(1).  
One way to obtain this result for $Z_N$ is directly from the recurrence 
relation of Eq.~(3).  The mean number of pions can be obtained from 
$\sum k x_k$ and is
$$
\left<N\right> = {ay \over 1-y} \eqno (35)
$$
The variance follows from $\left<N^2\right> - \left<N\right>^2 =
ay/(1-y)^2$ or
$$
\left<N^2\right> - \left<N\right>^2 = \left<N\right> \left(1 +
{\left<N\right> \over a}\right) \eqno (36)
$$
The result of Eq.~(36) is the result for the variance of a negative
binomial distribution.  This result is also similar to that of Eq.~(13)
when $a = 1$, but Eq.~(13) refers to fluctuations in a particular level $k$, 
while Eq.~(36) has no reference to a level.  The fluctuations can be much 
larger than Poissonian because of the factor $(1 + \left<N\right>/a)$.  To 
see that the probability distribution associated with this model
$\left(x_k = ay^k/k\right)$ is  indeed a negative binomial distribution,
the $P_N
= Z_N/z$ can be easily evaluated and rewritten as
$$
P_N = \left(N + a-1 \atop N\right)\,
\left({1 \over 1 + {\left<N\right> \over a}}\right)^a
\left({\left<N\right>/a \over 1 + {\left<N\right> \over a}}\right)^N
\eqno (37)
$$
The result of Eq.~(37) is the standard form of the negative binomial
distribution.

An interesting transformation of the negative binomial is to a set of clan
variables $N_c$, $n_c$, given by \cite{hov89}
\begin{eqnarray*}
N_c &=& a\log \left(1 + {\left<N\right> \over a}\right)\cr
n_c &=& {\left<N\right> \over N_c} &(38)\cr
\end{eqnarray*}
The $N_c$ equals the average number of clans, and $n_c$ is the average number
of particles in a clan.  The variables $N_c$, $n_c$ are a useful way of
analyzing experimental data.  In terms of $y$ and $z = 1/(1-y)^a$
\begin{eqnarray*}
N_c &=& \log z\cr
n_c &=& - {y \over (1-y) \log (1-y)} &(39)\cr
\end{eqnarray*}
Thus $n_c$ only depends on $y$ and $N_c$ is simply connected to $z$.  Note that
in thermodynamics (see also Sec.~II.B), the grand canonical partition function
$z$ is connected to the equation of state.  Specifically, $pV/k_B T = \log
z$ so that $pV = N_c k_B T$ if this connection is made.

\subsection{Other Models}

\nobreak
Besides the choice $\tau = 1$, other choices for $\tau$ can be solved.  Many of
these other choices were studied in the framework of cluster models 
\cite{cha94}, but the results obtained also apply to particle production of 
bosons.  Some discussion of these issues can be found in Ref. \cite{mek97} 
where results from a somewhat general theory of disordered systems were 
connected to distributions used in particle physics and quantum optics.

\section{CONCLUSIONS}

\nobreak
This paper discusses several models of pion production in a unified way by
showing that they are different choices of a quantity which we called
$x_k$.  The models considered are a thermal description of pion production, 
a relativistic hydrodynamic model, an emitting source pion laser model, and a 
model that gives rise to a negative binomial distribution.  The results of 
these models are compared with each other.  Comparisons are also made with 
Bose-Einstein condensation of Bose particles or atoms trapped in a harmonic 
oscillator well and in a box of $d$-dimensions.

The quantity $x_k$ appears in the weight given to each cycle class
decomposition of the symmetric group.  The index $k$ is the length of a cycle.
The cycle classes appear when the density matrix is symmetrized as in
Feynman's path integral approach to statistical mechanics.  The quantities
$x_k$, $k=1,2,3,\ldots$ completely determine the behavior of the system as
discussed in Sec.~II.A\null.  Specifically the mean number of pions and their
fluctuations are just moments of the $x_k$ distribution.  The grand canonical
partition function $z$ is an also simply obtained from these $x_k$'s, and the
canonical partition function $Z_N$ can be obtained by a simple recursive
procedure which contains the $x_k$.  The probability of $N$-pions is $P_N =
Z_N/z$.  These methods and relations were initially used by one of us in
cluster problems \cite{mek90}--\cite{cha94} and subsequently applied to 
Bose-Einstein cycle class problems \cite{cha94,chapre}.  They were also used 
in a discussion of the pion laser model in Refs. \cite{pra93}--\cite{cso77}.  
The two problems, cycle class and clusters, can be mapped into one another as 
shown in Fig.~1.  Specifically the index $k$ is either the cluster size or 
cycle length, and $n_k$ is either the number of cycles of length $k$ or the 
number of clusters with $k$ nucleons.  After presenting various general 
expressions, specific cases are discussed and compared.  We started with the 
thermal model because of its long history and its widespread use in heavy ion 
collisions.The thermal model was used to motivate some of the quantities, such 
as the $x_k$'s that also appear in the other cases that are considered.  The 
role of the Boltzmann factor $e^{-km/T}$ in $x_k$, which represents the cost 
function for making $k$-pions, is important in the thermal model.  The mean 
number of pions, their fluctuations, and the probability distribution for 
$N$-pions are discussed and compared with Poisson results.  The results for a 
relativistic hydrodynamic model are also discussed.  The relativistic 
hydrodynamic model is used in the analysis of CERN/NA44 and CERN/NA49 data.  
The probability distribution for thermal pions is shown to be a Gaussian 
distribution.  The fluctuations in pions is about 20\% above the Poisson limit.
This result of slightly enhanced fluctuations is also consistent with thermal 
model results.

The emitting source model of S.~Pratt is next discussed.  It has an interesting
feature of the possibility of large departures from Poissonian results.  This
model also has some other interesting properties that are not present in the
thermal and hydrodynamic models.  Namely, a critical pion density $\eta_c$
exists and, if the Poisson emission strength $\eta$ equals this density, an
infinite number of pions results from Bose-Einstein enhancement factors.  The
thermal and hydrodynamic models lack this property because $\eta \sim
\eta_c e^{-m/T}$.  Some result of the emitting source model are shown to be
similar to Bose-Einstein condensation in a laser trap when this trap is
taken to be a three-dimensional harmonic oscillator well.  Another 
distribution which can have large fluctuations above Poisson results is the 
negative binomial distribution.  The clan representation of the negative 
binomial distribution, introduced by Van~Hove and Giovannini, is discussed in 
terms of quantities that appear in the specification of the $x_k$'s.  The clan 
representation is a useful way of characterizing data, and it is compared with 
the thermal model results.

\bigskip

This work has been supported by the U.S. Department of Energy.

 \begin{table}
 \caption{The probability $P_N$ of $N$ pions compared to a Poisson
distribution with the same $\left<N\right>$.  Also given are the cycle class 
parameters $x_k$ obtained from Eq.~(14).}
 \label{table1}
 \begin{tabular}{cllcr}
$N$ & $P_N$ & Poisson & $k$ & $x_k$\\
\hline
0& .311& .293&&\\
1& .346& .359& 1& 1.10812\\
2& .208& .221& 2& .04812\\
3& .0903& .0904& 3& .00588\\
4& .0320& .0278& 4& .00107\\
5& .00998& .00682& 5& .00024\\
6& .00287& .00139& 6& .00006\\
 \end{tabular}
 \end{table}

\begin{table}
\caption{The distribution of $x_k$ in a Relativistic Hydrodynamic Model}
 \label{table2}
\begin{tabular}{crlr}
$k$ & $x_k$ & $k$& $x_k$\\
\hline
1& 260& 6& .0067\\
2& 9.957& 7& .0016\\
3& 1.047& 8& .0004\\
4& .163& 9& .0001\\
5& .031& & \\
 \end{tabular}
 \end{table}

\begin{table}
\caption{Values of $x_k$.  Unless $\eta$ is large, the $x_k$ fall very
rapidly and thus the $\left<N\right>$ and $\left<N^2\right> - 
\left<N\right>^2$ are governed by the first few $k$'s.  For the coice 
$R^2/2 = 6$ and $1/8 mT = 1/6$, the $\eta_c = 42.875$.  However, the second 
$x_k = x_2$ is comparable to $x_1$ at $\eta = 432$, the $x_3$ to $x_2$ at 
$\eta = 140$, the $x_4$ to $x_3$ at $\eta = 90$, etc.}
 \label{table3}
\begin{tabular}{cccc}
$k$ & $x_k$ & $k$& $x_k$\\
\hline
1& $\eta$& 6& $\eta^6/2.43 \times 10^{10}$\\
2& $\eta^2/432$& 7& $\eta^7/1.32 \times 10^{12}$\\
3& $\eta^3/6.07 \times 10^4$& 8& $\eta^8/7.40 \times 10^{13}$\\
4& $\eta^4/5.47 \times 10^6$& 9& $\eta^9/3.8 \times 10^{15}$\\
5& $\eta^5/3.91 \times 10^8$& 10 &$\eta^{10}/2.07 \times 10^{17}$ \\
 \end{tabular}
 \end{table}

\begin{table}
\caption{Values of $x_k$ in a thermal model using the same choice of $R^2$, $T$
as in Table~2.  The $x_k$ is also written as $\eta^k/b_k$ where $x_1 =
\eta$ and $b_k$ is a number that is given in the table.  The results 
$\eta^k/b_k$ are very similar to that of Table~3.}
 \label{table4}
\begin{tabular}{ccc}
$k$ & $x_k$ & $x_k = \eta^k/b_k$\\
\hline
1& 20& $\eta$ \\
2& 0.991& $\eta^2/404$\\
3& 0.142& $\eta^3/6.21 \times 10^4$\\
4& $0.309 \times 10^{-1}$& $\eta^4/5.17 \times 10^6$\\
5& $00.835 \times 10^{-2}$& $\eta^5/3.83 \times 10^8$\\
6& $00.257 \times 10^{-2}$& $\eta^6/2.57 \times 10^{10}$\\
7& $000.869 \times 10^{-3}$& $\eta^7/1.47 \times 10^{12}$\\
8& $000.312 \times 10^{-3}$& $\eta^8/8.20 \times 10^{13}$\\
9& $000.118 \times 10^{-3}$& $\eta^9/4.34 \times 10^{15}$\\
10& $0000.462 \times 10^{-4}$& $\eta^{10}/2.22 \times 10^{17}$\\
 \end{tabular}
 \end{table}

\begin{figure}
\caption{Cycles and Clusters.  In the mapping of a cycle class problem into 
a cluster problem and vice versa, the cycle length corresponds to the cluster 
size; the number of cycles of length $k$ corresponds to the number of 
clusters with $k$ nucleons.  Also shown is the vector $\vec n = (n_1, n_2, 
\ldots, n_A)$ as a block diagram.  The number of vertical blocks is $k$, and 
the number of columns with $k$ blocks is $n_k$.  The total number of blocks 
is $A$.}
\end{figure}

\begin{figure}
\caption{Various Parallels.  Various parallels are shown which include the
partitioning of an integer, a corresponding block picture, a cluster,  a
cycle class representation of each partition, and an equivalent harmonic
oscillator excitation.  Some places where these various columns appear are
as follows:  The partitions of an integer appears in counting the number of
irreducible representation of the symmetric group.  Block pictures and
rotated versions of the ones shown appear in combinatorial analysis (Ferrer's 
block diagram) and in group theory (Young tableaus).  Cluster problems appear 
in many areas of physics such as nuclear multifragmentation, percolation, 
randomly broken objects, and the fragmentation of atomic clusters.  They also 
appear in group structure such as in the distribution of city sizes, etc.  
Cycle class problems appear in Bose-Einstein and Fermi-Dirac statistics,
speckle patterns in phase space, Feynman's theory of the $\lambda$-transition 
in liquid He, in random permutations, etc.  Excitations in the harmonic 
oscillator appear when evaluating the microcanonical partition function of 
Bose atoms in a laser trap. Another application is to the Veneziano-Hagedorn 
mass spectrum, where each excitation is an elementary particle.}
\end{figure}

\begin{figure}
\caption{Results for $P_N$ for thermal pions, $\pi^-$, in 158 $A\cdot GeV$
Pb+Pb collisions. The solid line uses contributions $k \rightarrow 600$
(already $k=10$ gives the fully converged result). The dotted lines only uses
up to $k=2$ terms. In the figure, we have (solid line) $\Sigma x_k = 271.2$,
$\langle N \rangle = \Sigma k x_k = 283.9$,
and $\langle N^2 \rangle - \langle N \rangle^2  = \Sigma k^2 x_k = 312.97$.}
\end{figure}

\clearpage
{\huge \bf Figure 1:}
\null
\vspace*{-19.0cm}
\begin{figure}
\begin{center}\mbox{ }
\[
\psfig{figure=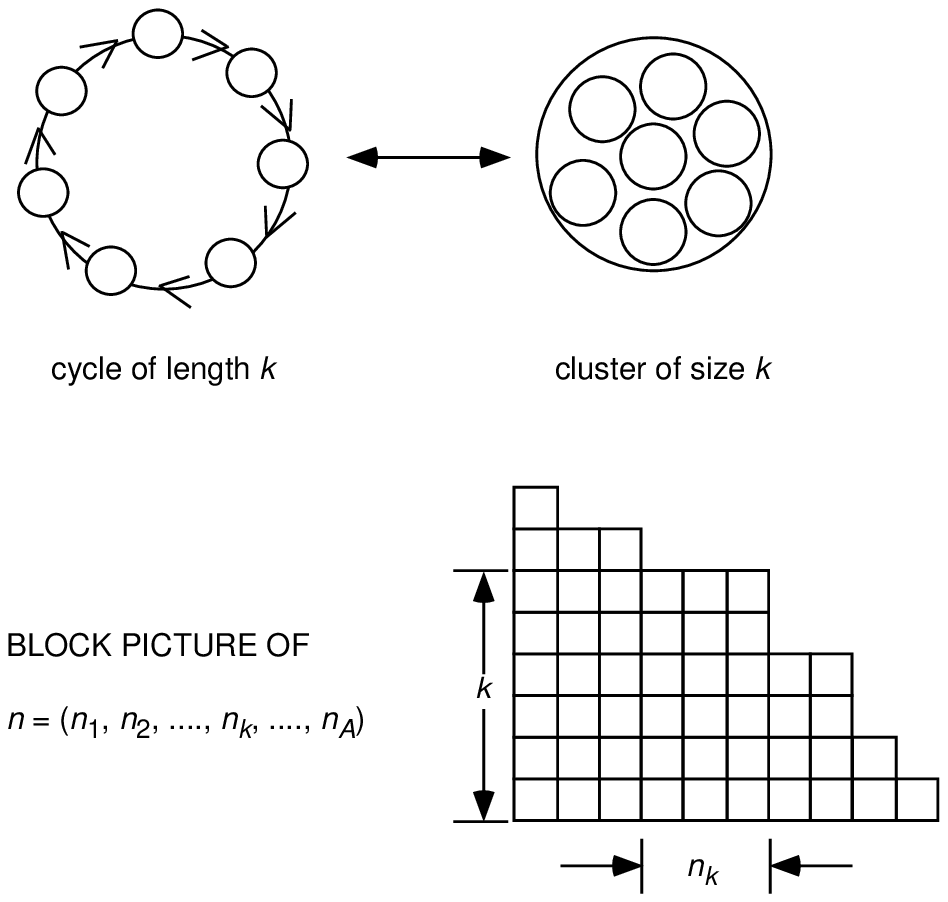,bbllx=0.0cm,bblly=10.0cm,%
bburx=21.0cm,bbury=29.7cm,width=21.0cm}
\]
\end{center}
\label{fg:fig1}
\end{figure}

\hspace*{0.0cm}
\vspace*{4.5cm}
\begin{figure}
\begin{center}\mbox{ }
\[
\psfig{figure=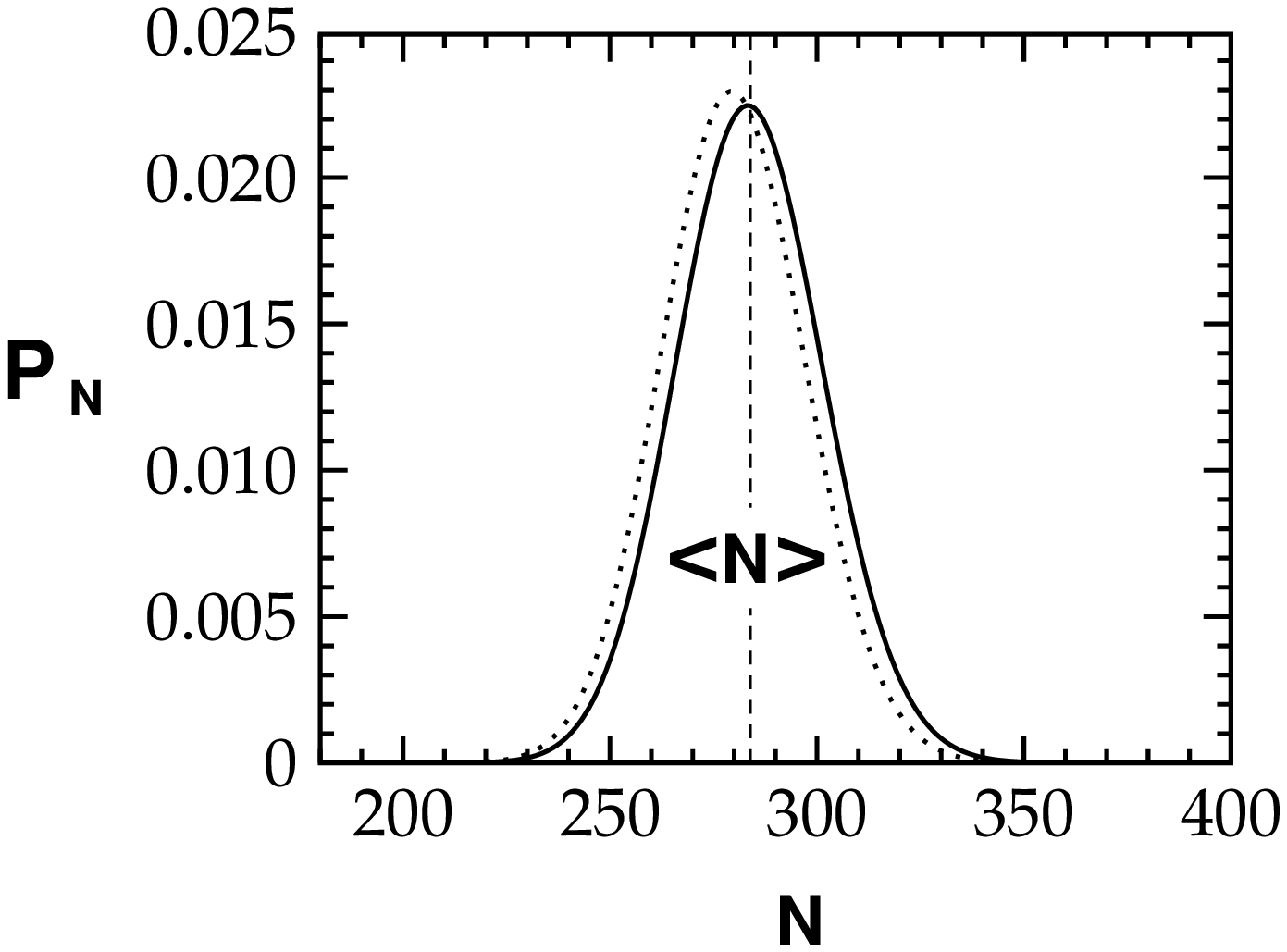,bbllx=3.0cm,bblly=15.0cm,%
bburx=21.0cm,bbury=27.7cm,width=18.0cm,clip=}
\]
\end{center}
\vspace*{-12.5cm}
{\huge \bf Figure 3:}
\label{fg:fig3}
\end{figure}

\clearpage
{\huge \bf Figure 2:}
\null
\vspace*{-9.5cm}
\begin{figure}
\begin{center}\mbox{ }
\[
\psfig{figure=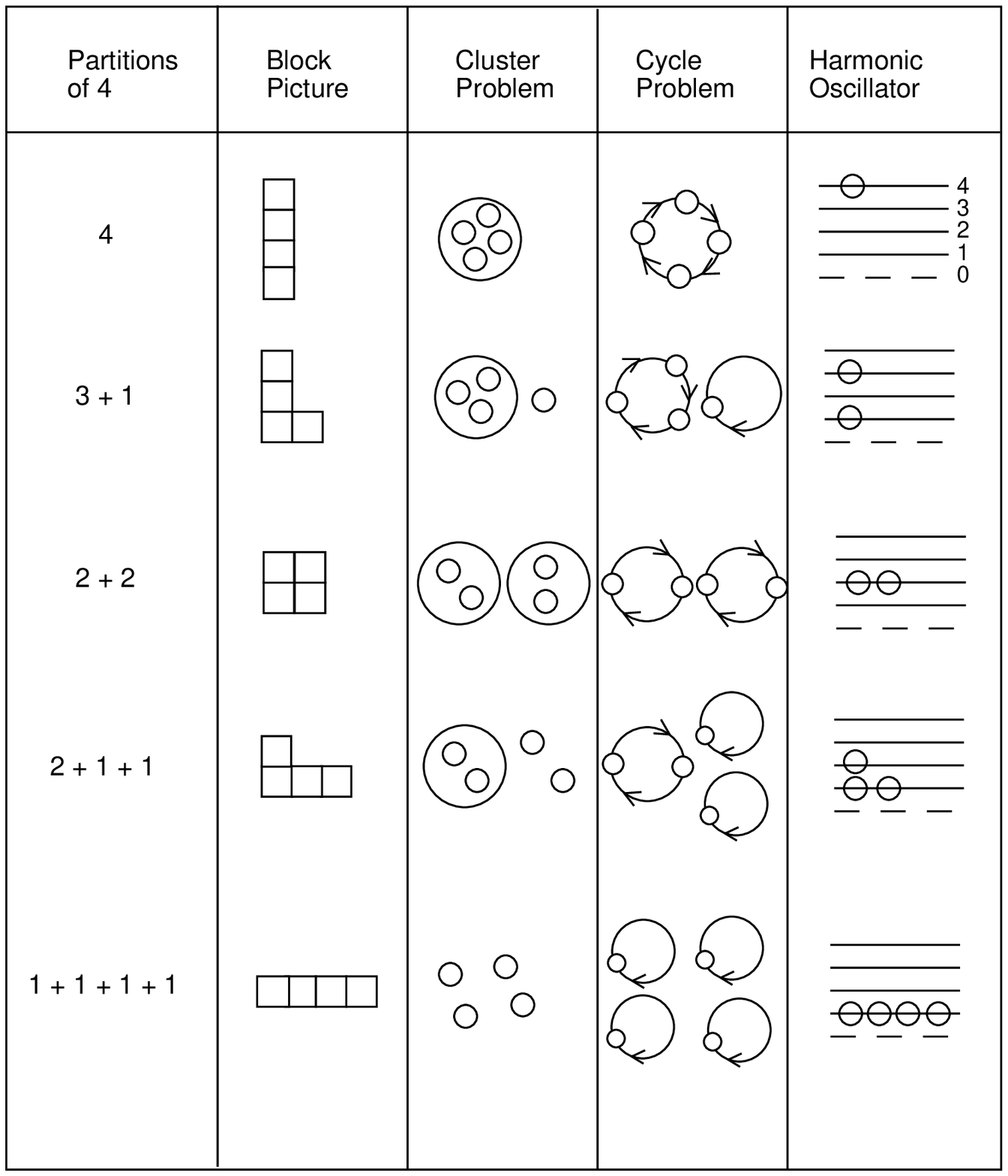,bbllx=0.0cm,bblly=10.0cm,%
bburx=21.0cm,bbury=29.7cm,width=21.0cm}
\]
\end{center}
\label{fg:fig2}
\end{figure}

\end{document}